\documentclass[aps,prc,twocolumn,floatfix,superscriptaddress]{revtex4}
\usepackage{epsfig}
\usepackage{amsmath}
\usepackage{color}
\usepackage[utf8]{inputenc}
\usepackage{graphicx}

\definecolor{blue}{rgb}{0.05, 0.05, 0.5}

\begin{document}

\title{Ratio of photon anisotropic flow in relativistic heavy ion collisions}

\author{Rupa Chatterjee}
\email{rupa@vecc.gov.in}
\affiliation{Variable Energy Cyclotron Centre, HBNI, 1/AF, Bidhan Nagar, Kolkata 700064, India}
\author{Pingal Dasgupta}
\email{pingaldg@fudan.edu.cn}
\affiliation{Key Laboratory of Nuclear Physics and Ion-beam Application (MOE), Institute of Modern Physics, Fudan University, Shanghai 200433, China}

\begin{abstract}
The $p_T$ dependent  elliptic and triangular flow parameters of direct photons are known to be dominated by thermal radiations. The non-thermal  contributions dilute the  photon anisotropic flow  by adding extra weight factor  in the $v_n$ calculation.
The discrepancy between experimental photon anisotropic flow data and results from theoretical model calculations is  not well understood even after significant developments in the model calculations as well as in the experimental analysis.
We show that the ratio of photon $v_n$  can be a potential observable in this regard by minimizing the uncertainties arising due to the non-thermal contributions.
 We calculate the $v_2/v_3$ of photons as a function of $p_T$ from heavy ion collisions  at RHIC and compare the results with available experimental data.
The ratio does not change significantly  $p_T$ in the region $p_T>2$  GeV. However, it rises towards smaller $p_T$ ($< 2$ GeV) values. The 
ratio is found to be  larger for peripheral collisions than for central collisions. In addition, it is found to be  sensitive to the initial formation time and the final freeze-out temperature at different $p_T$ regions unlike the individual anisotropic flow parameters. We  show that the photon $v_1/v_2$ and $v_1/v_3$ along with the $v_2/v_3$ results may help us constraining the initial conditions.
\end{abstract}

\pacs{25.75.-q,12.38.Mh}

\maketitle

\maketitle
The anisotropic flow of photons produced in relativistic heavy ion collisions can be a potential observable to study the initial state and the evolution of the hot and dense medium formed in such collisions~\cite{Srivastava:2008es, Chatterjee:2005de, Gale:2014dfa, Chatterjee:2013naa, Monnai:2014kqa, McLerran:2014hza, Basar:2012bp,Tuchin:2012mf,Zakharov:2016mmc,rc_pramana,rc_cluster}. 
The main advantage of studying direct photon observables in heavy ion collisions is that their emission is quite sensitive to the initial state of the produced matter~\cite{vesa}. As a result, even  a slight change in the initial conditions would result in significant change in the final state photon observables~\cite{Vujanovic:2014xva,Liu:2012ax,rc_tau0,rc_phot_fluc, Chatterjee:2011rg,Dasgupta:2017fns}. 

The study of photon anisotropic flow has become a topic of great interest in the heavy ion community in last couple of decades.  This is due to the unique $p_T$ dependent nature of the flow parameter compared to the anisotropic flow of hadrons and most  importantly due to the persistent tension between experimental photon $v_2$ data~\cite{Adare:2008ab,Adam:2015lda} and the results from theory calculations. 
The elliptic as well as triangular flow of photons show interesting features as a function of the transverse momentum due to the competing contributions from the hot and dense plasma and relatively cold hadronic matter phases~\cite{rc_v3,rc_v2_v3}. The main concern still remaining  is that the theoretical model calculations underpredict the experimental direct photon $v_n$  data both at RHIC and LHC energies by a significant margin which is called the direct photon puzzle~\cite{Adare:2015lcd,Acharya:2018bdy}.


There has been significant development in the photon $v_n$ calculation since the initial predictions  of thermal photon anisotropic flow using a hydrodynamical model framework~\cite{Chatterjee:2005de} to understand the puzzle better.  The presence of fluctuations in the initial density distribution is found to increase the elliptic flow compared to a smooth initial density distribution and also gives rise to non-zero triangular flow of photons. The photon anisotropic flow from Cu+Cu collisions is calculated to be larger than from Au+Au collisions at a particular centrality bin as the initial state fluctuations are found to be more effective for the smaller systems. On the other hand, the inclusion of shear viscosity in the hydrodynamical model calculation  is found to reduce the elliptic flow especially towards the higher $p_T$ values. It has also been shown recently that  the directed flow parameter $v_1$ of photons  can play a crucial role in  our understanding of photon observables in heavy ion collisions~\cite{Shen:2013cca,rc_v1}. The  photon $v_1 $  is found to show  distinct $p_T$ dependent behaviour compared to the elliptic and the triangular flow parameters. The $v_1$ is found to be negative for smaller $p_T$ and becomes positive at higher $p_T \ (> 2 \ {\rm GeV})$ values. In addition, the  $v_1$ is found to be completely dominated by the QGP radiation~\cite{rc_v1}.

The direct photon measurement from a number of systems at wide range of beam energies has also been reported in recent times. 
 Low $p_T$ direct photons have been measured by the PHENIX Collaboration from Cu+Cu collisions at 200A GeV a RHIC for 0-40\% as well as for minimum bias configurations~\cite{phenix_cucu,Khachatryan:2018uzc}. An excess of direct photon yield over the pp baseline has been reported for Cu+Cu collisions and the photon data is also found to be close to the Au+Au results at similar ${\rm N_{part}}$ values. 
The direct photon measurements at 39 and 62.4 GeV Au+Au collisions at RHIC by the PHENIX Collaboration~\cite{phenix_39_62.4} also reported  a significant enhancement in photon yield compared to the pp baseline  at these lower beam energies.
In addition, a scaling behaviour has been observed in the experimental data where the $dN_{\gamma}^{\rm {dir}}/d\eta$ is found to scale with the charged particle multiplicity $(dN_{ch}/d\eta)^\alpha$ where, the value of $\alpha$ is found to be about 1.25 in the region $1\le p_T \le 5$ GeV~\cite{Khachatryan:2018uzc}. 

These recent developments in the  experimental analysis  hold the promise that in near future photon anisotropic flow data will be available from a number of  systems at different beam energies which will add up to our understanding of initial state as well as photon anisotropic flow from heavy ion collisions~\cite{rupa,gabor}. 



\section{ratio of photon anisotropic flow}
We know that thermal photons produced from the QGP phase and hot hadronic matter dominate the  photon anisotropic flow as the contributions from all other sources of direct photon are negligible compared to the thermal radiations. The QGP contribution decides the $p_T$ dependent shape of the photon anisotropic flow, whereas the thermal photons from hot hadronic matter decides the magnitude of the photon $v_n$. We see larger photon $v_2$ for  peripheral collisions as the relative contribution from the hadronic phase (and also the initial spatial eccentricity) increases for more glancing collisions. The non-thermal contributions (mostly the prompt photons) indirectly affect  the photon $v_n$  by adding extra weight in the denominator of the photon anisotropic flow calculation.

The initial conditions of the model calculation play crucial role in the determination of the photon anisotropic flow in heavy ion collisions. Photon observable are more sensitive to  the formation time ($\tau_0$) and the initial temperature compared to  hadronic observables. A smaller $\tau_0$ leads to larger initial temperature and relatively larger contribution from the QGP phase (compared to the photons produced in hadronic matter). This leads to smaller photon anisotropic flow at larger $p_T$~\cite{rc_tau0}. Thus, initially it was assumed that photon anisotropic flow can be a useful observable for precise determination of $\tau_0$ value. However, the discrepancy between the  data and model calculation has completely ruled out this possibility.


We suggest that the ratio of photon $v_2$ and $v_3$ as a function of $p_T$ can be a useful parameter in this regard.  We show that one can reduce the uncertainties arising due to the non-thermal contributions as well as  constrain the initial conditions by taking ratio of the  anisotropic flow parameters.  A comparison with the experimental data  also provide valuable information about the initial conditions.

The anisotropic flow of direct photons can be approximated interms of thermal and non-thermal contributions as
\begin{equation}
v_n \ = \ \frac{v_n^{\rm {th}} \ \times \ {\rm dN^{th}}} { \rm{dN^{th} \ + \ dN^{non-th}}}
\end{equation}
where, we have neglected the very small contribution of $v_n^{\rm non-th}$.

The denominator or the total (thermal $+$ non-thermal) photon yield  for a particular $p_T$ remains same for $v_2$, $v_3$ and also for $v_1$. The ratio of $v_2/v_3$ consists of thermal contribution only by cancelling out the total photon yield from the denominator of individual $v_2$ and $v_3$ results. 
Thus, the ratio  can be a relatively cleaner observable compared to the photon $v_n$.

We study the ratio of thermal photon $v_2$ and $v_3$ from Au+Au, Cu+Au, and Cu+Cu collisions at RHIC and compare it with available experimental data. The ratio $v_1/v_2$ (and $v_1/v_3$) is also calculated at RHIC. The sensitivity of the results to the initial conditions are checked.
 
We use an event-by-event ideal hydrodynamical framework for this study. 
 The  standard Woods-Saxon nuclear density distribution with a  Monte Carlo Glauber model is used for the initial  energy density distribution. The initial parameters for the model calculation are fixed by reproducing the charged particle multiplicity, spectra and anisotropic flow parameters at RHIC.
The $\tau_0$ for  Au+Au collisions at RHIC is taken as 0.17 fm/$c$ and the freeze-out temperature as 160 MeV. 
 The photon rates from QGP and hadronic matter  are taken from Refs.~\cite{amy,nlo_thermal,trg}. 

\begin{figure}[h]
\centerline{\includegraphics*[width=8.4 cm]{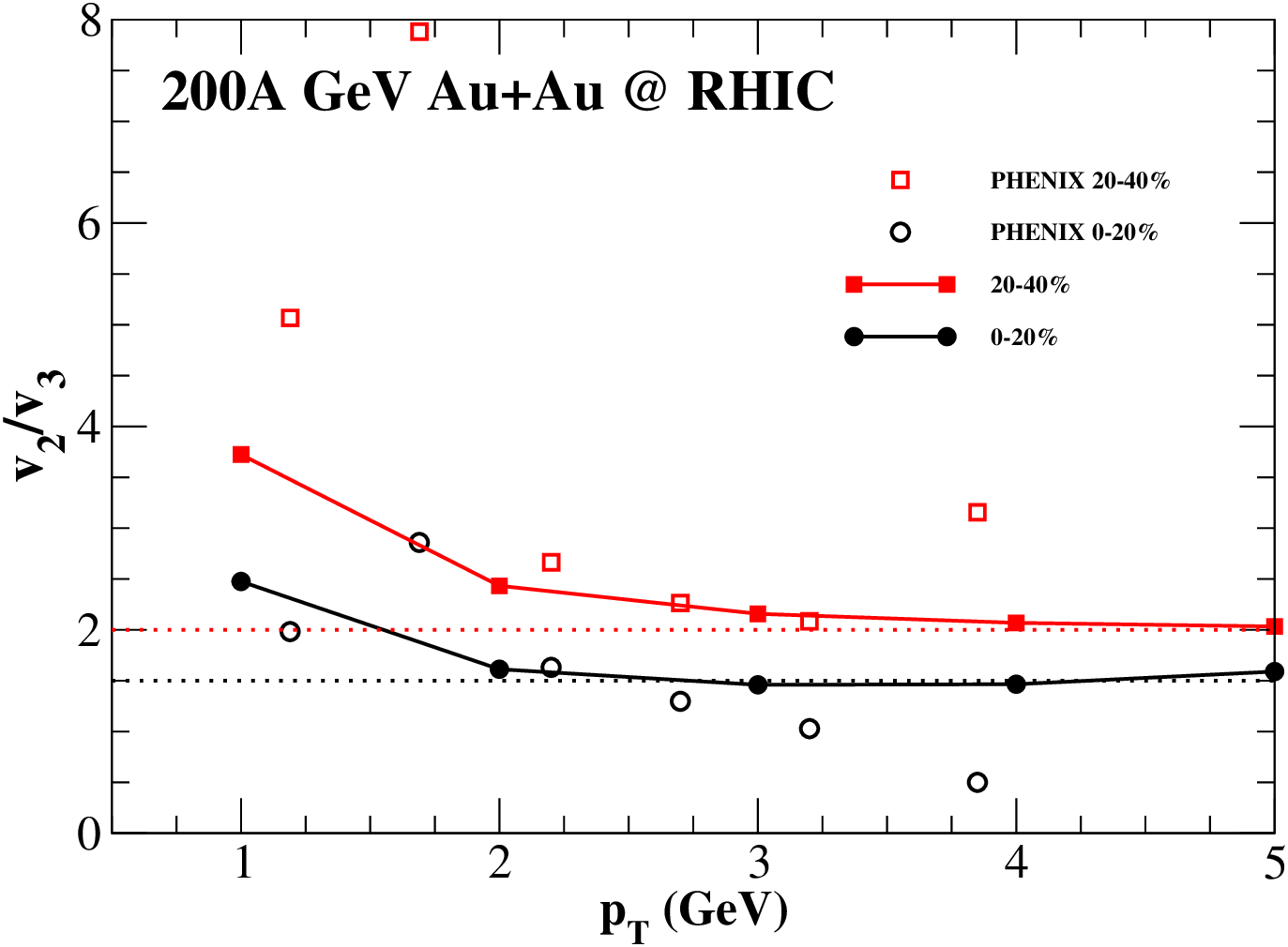}}
\caption{(Color online) The ratio of thermal photon $v_2$ and $v_3$ as a function of $p_T$ from Au+Au collisions at RHIC for 0--20\% and 20--40\% centrality bins.}
\label{fig1}
\end{figure}

\section{results}

In Fig.~\ref{fig1} we show the ratio of thermal photon anisotropic flow ($v_2/v_3$) from two different centrality bins of Au+Au collisions at RHIC. The ratio of PHENIX direct photon data~\cite{Adare:2015lcd} is  plotted on the same figure for comparison. The error bars are avoided in the plot as uncertainties in $v_2/ v_3$ data is not published, however, it is expected to be a large quantity. Precise measurements of the observable in the near future are expected to reduce the error bar.  
 The individual $v_2$ and $v_3$ results from hydrodynamical model calculations under-estimate the data by a large margin. However, the ratio of $v_2/v_3$ from both experimental data and model calculations is found to be close to each other in the region 2--4 GeV $p_T$. This $p_T$ range is expected to be dominated 
by the QGP radiation in the direct photon spectrum.

We have shown earlier  that the ratio of the elliptic and the triangular flow  of thermal photons shows interesting features both at RHIC and at the LHC~\cite{rc_v2_v3}. The ratio does not change significantly in the region $p_T > $ 2 GeV whereas, it rises with smaller values of $p_T \ (< \ 2 \ {\rm {GeV}})$.
These results shows that the $p_T$ dependent behaviour of $v_2$ and $v_3$ are  close to each other in the higher $p_T$ region as the ratio remains almost constant.


The ratio of $v_2/v_3$ calculated from Pb+Pb collisions  at LHC also shows similar features~\cite{rc_v2_v3}.
At LHC,  the QGP phase lived longer than at RHIC for a particular centrality bin. Probably due to this reason the ratio is seen to be almost  independent of the value of transverse momentum upto a larger $p_T$ value  at the  LHC than at RHIC.

We have checked that an event-by-event distribution of photon $v_2$ and $v_3$ (at different $p_T$ values) does  not show any correlation between them. However, the  event averaged ratio between them  shows different nature than the event-by-events results specially at higher $p_T$ values. The ratio is always greater than 1 as the elliptic flow is larger than the triangular flow parameter. In addition, the ratio increases for peripheral collisions as photon $v_2$ shows stronger sensitivity to the collision centrality compared to the triangular flow parameter. 



\begin{figure}[h]
\centerline{\includegraphics*[width=8.4 cm]{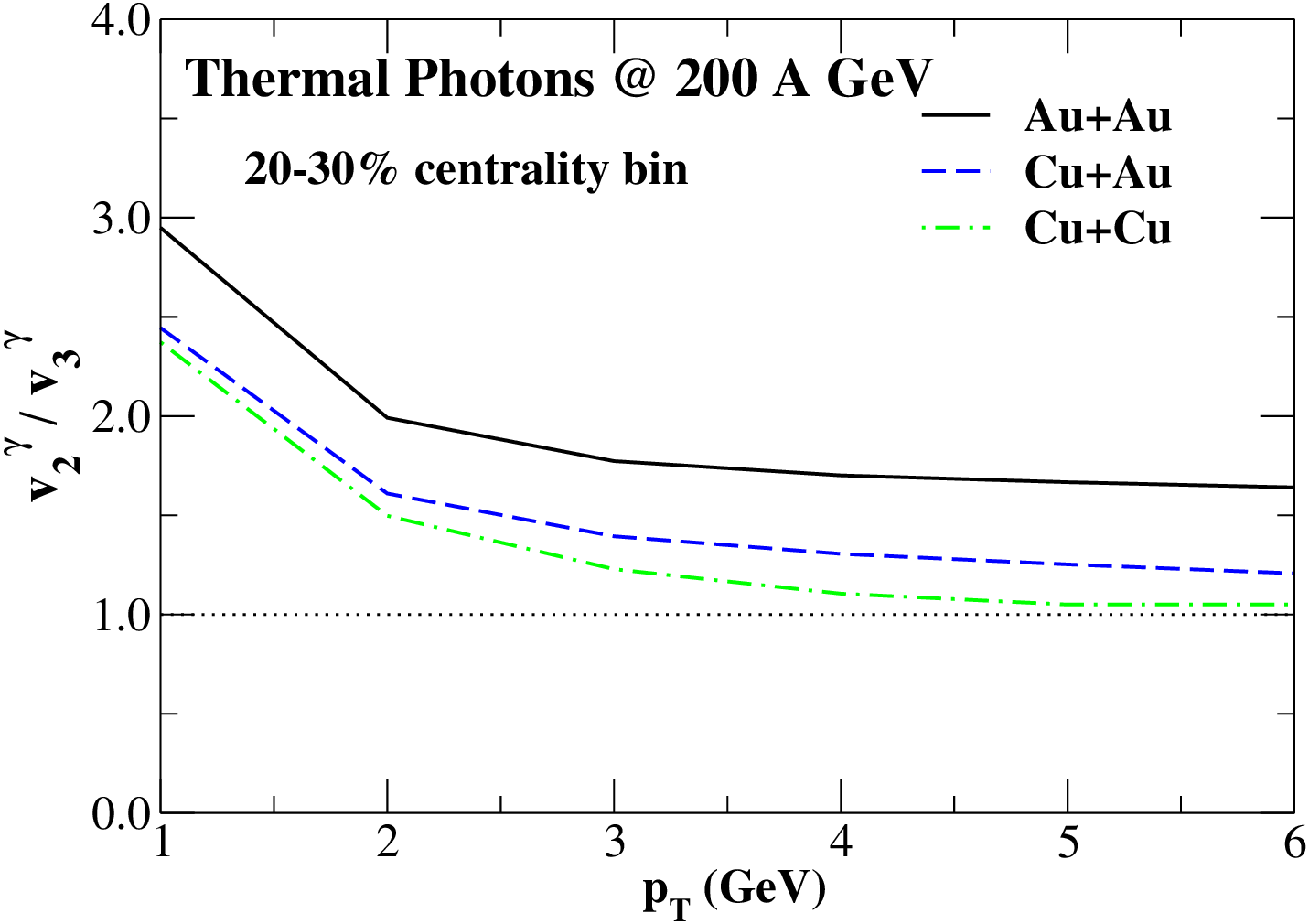}}
\caption{(Color online) The ratio of thermal photon $v_2$ and $v_3$ as a function of $p_T$ from Au+Au, Cu+Au and Cu+Cu collisions for centrality bin 20--30\% at RHIC. }
\label{ratio_20_30}
\end{figure}

The $v_2/v_3$ results from  three different colliding systems at RHIC  are shown in Fig.~\ref{ratio_20_30}. The ratio shows similar qualitative nature for all three cases and 
 is found to be  largest for the Au+Au collisions. The results from Cu+Au and Cu+Cu collisions are found to close to each other. In addition,  we do not see any significant  difference in the  $p_T$ dependent nature between the symmetric (Au+Au, Cu+Cu) and asymmetric (Cu+Au) collisions. The effect of initial state fluctuations is  more for smaller systems which reduces the gap between  $v_2$ and $v_3$ and consequently a relatively smaller value of the ratio can be seen for those. 

Photon production from the hadronic matter is significant in the direct photon spectrum for $p_T <$ 2 GeV where the ratio rises with  $p_T$. 
 However, we know that the photon $v_2$  at higher $p_T$ values are larger than the (only) QGP $v_2$ due to the presence of small (compared to QGP) but effective contribution from the hadronic phase. Thus a $p_T$ dependent temporal build-up of the $v_n$ from the QGP and hadronic matter  would be useful to understand the ratio better and we discus that in the next section.

\begin{figure}[h]
\centerline{\includegraphics*[width=8.4 cm]{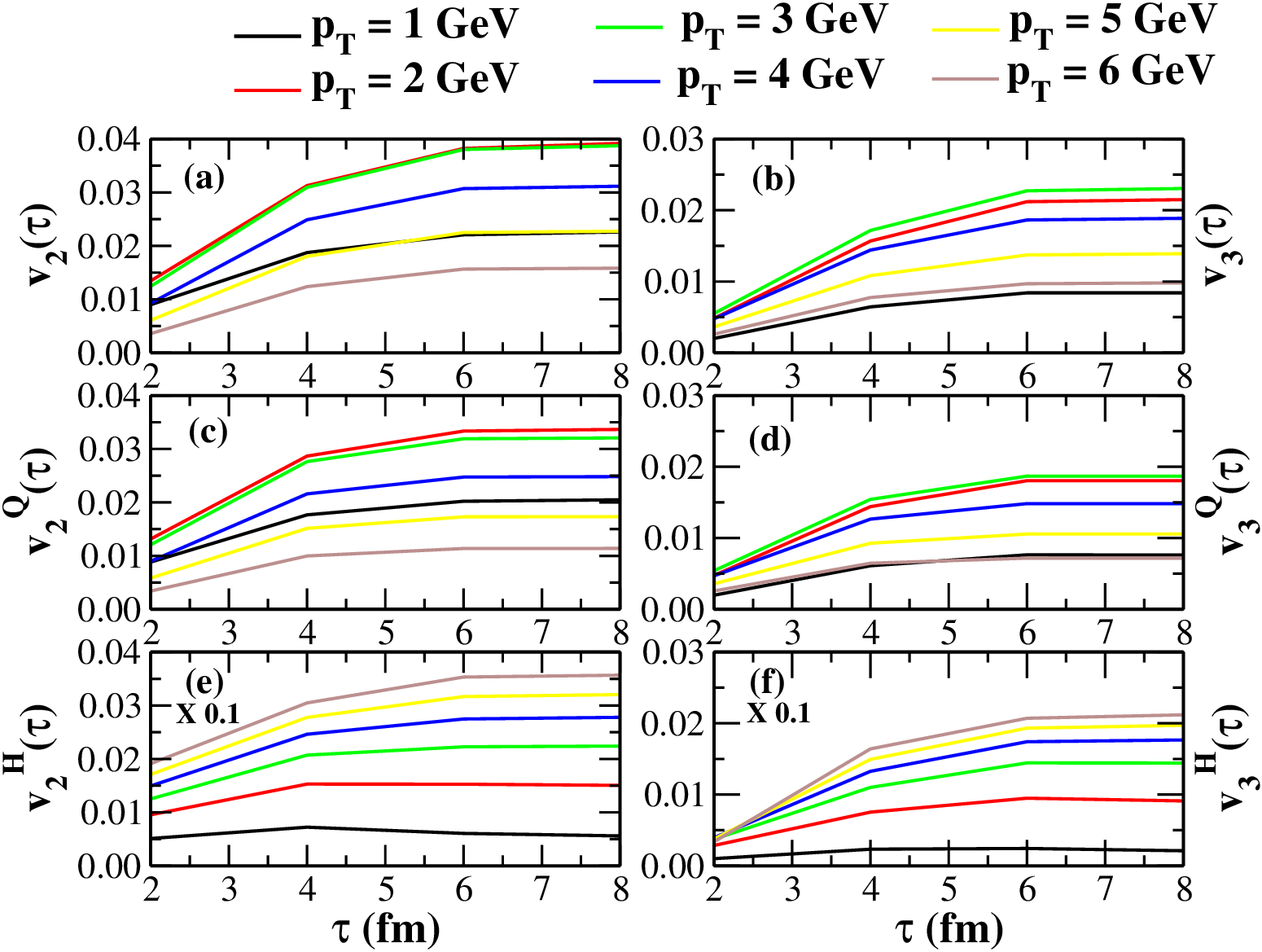}}
\caption{(Color online) Time evolution of (a) elliptic  and (b) triangular  flow parameter of thermal photons  at 6 different $p_T$ values from Au+Au collisions at RHIC and for 20--40\% centrality bin. Individual component of $v_2$ and $v_3$, (c)--(d)  from QGP  ( $v_2^Q$ and  $v_3^Q$) and (e)--(f) hadronic matter ($v_2^H$ and $v_3^H$) are shown in the subsequent plots. }
\label{ratio_auau}
\end{figure}
\begin{figure}[h]
\centerline{\includegraphics*[width=8.0 cm]{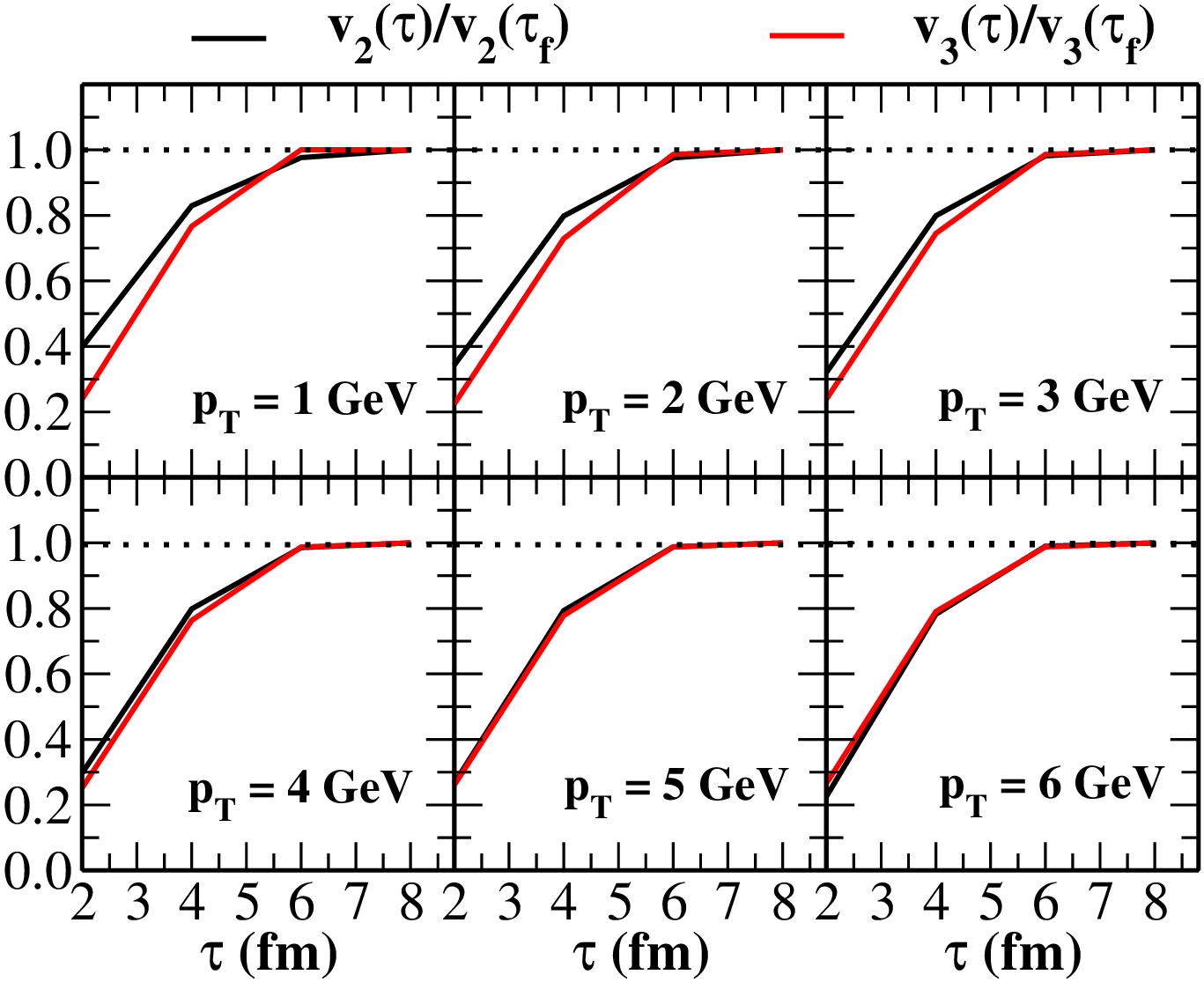}}
\caption{(Color online)  Time evolution of normalized total photon $v_2$ and $v_3$ at different $p_T$.}
\label{t}
\end{figure}
\section{Time evolution of photon anisotropic flow}
The time evolution (event averaged) of photon elliptic and triangular flow parameters at different $p_T$ values from Au+Au collisions at RHIC are shown in Fig.~\ref{ratio_auau}.  The top panels show the  total thermal  photon $v_n$ as a function of $\tau$, whereas the bottom two panels show results from (only) QGP and (only) hadronic matter respectively for 20--40\% centrality bin. 

The time evolution of (total) photon $v_n$ and the  (only) QGP $v_n$ shows similar qualitative behaviour. The maximum contribution to $v_2$ comes from the 2 - 3 GeV $p_T$ region in the QGP phase and the same is reflected in the total photon $v_2$ . Whereas, the hadronic contribution is largest for the 5 - 6 GeV $p_T$ values. In addition, the the photon $v_n$ can be seen to rise rapidly till 5 - 6 fm/$c$ time period and beyond that the its growth is seized. 


The time evolution of the $v_2$ and $v_3$ results are compared in Fig.~\ref{t} 
which explains the $p_T$ dependent ratio of photon $v_n$ clearly. The $v_n(\tau)$ at different $p_T$ values is normalized by the final photon $v_n(\tau_f)$ at those $p_T$ values. The photon $v_2$ at smaller $p_T$  develops little earlier than the triangular flow parameter  resulting in a difference between the two results. However, the for $p_T \ >$ 3 GeV, the elliptic as well as triangular flow parameters develops in a similar rate and thus the results are found to be quite close to each other. As a result at lower $p_T$ we see the ratio is larger and  it does not change significantly with rise in $p_T$.



\section{Ratio with directed flow}
\begin{figure}[h]
\centerline{\includegraphics*[width=8.4 cm, clip=true]{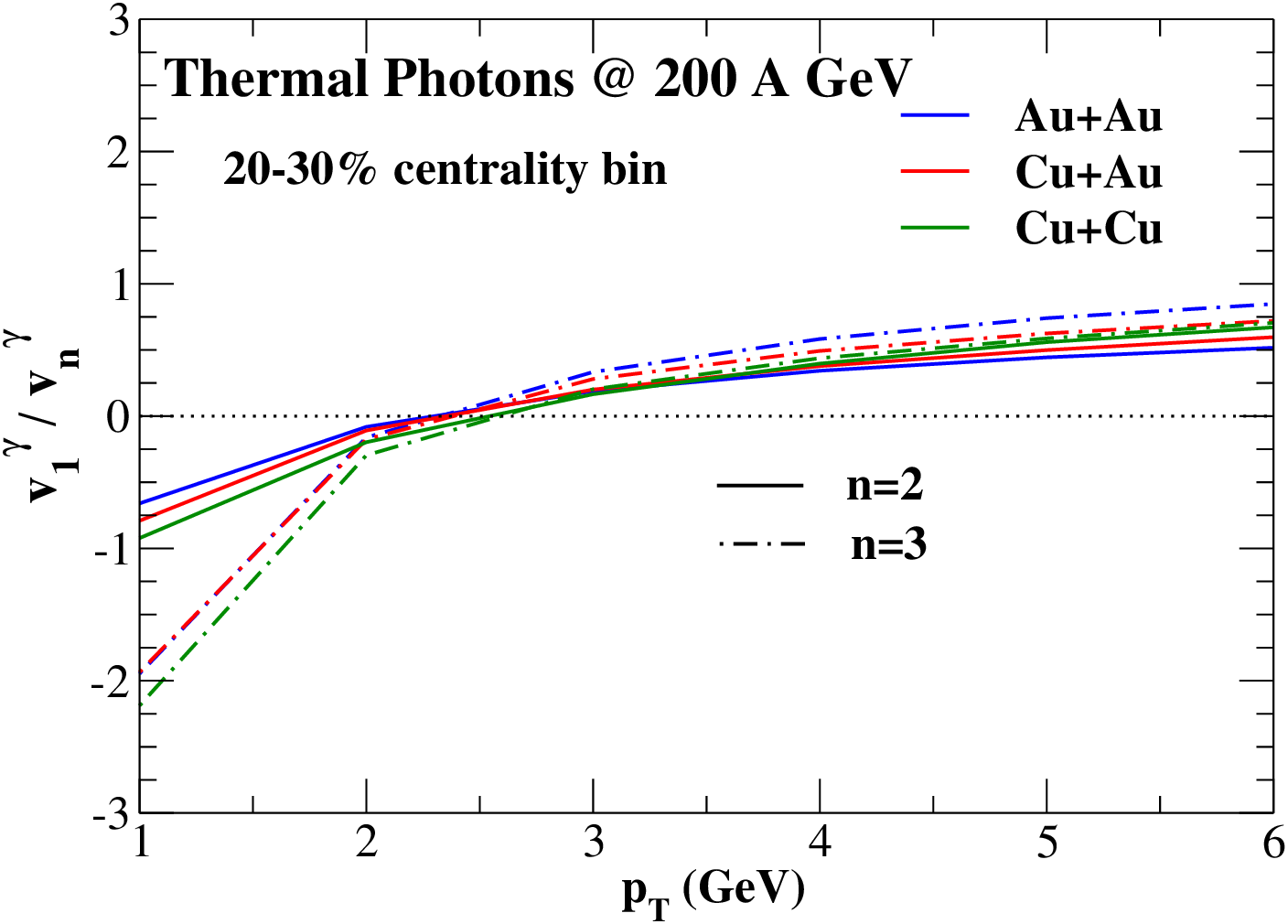}}
\caption{(Color online) The ratio thermal photon $v_n$  ($n=2,3$) and $v_1$  from Au+Au, Cu+Au and Cu+Cu collisions at RHIC and for 20--30\% centrality bin. }
\label{ratio_v1}
\end{figure}
The directed flow of photons has shown unique nature as a function of $p_T$ which is different from the elliptic and triangular flow parameters. It has been shown in Ref.~\cite{Shen:2013cca} that photon $v_1$ does not depend on the centrality of the collisions. We have also  seen that the directed flow of photons is totally dominated by the QGP radiations~\cite{rc_v1}. Thus, the experimental determination of photon $v_1$ is expected to provide valuable information regarding photon anisotropic flow parameter.

Fig.~\ref{ratio_v1} shows the ratio of directed and elliptic (triangular) flow parameters of photons as function of $p_T$ for three different symmetric as well as asymmetric collisions. 
 We see both positive and negative values of the ratio which does not show significant difference between the results from three different systems. The value of $v_1/v_2$  from Au+Au collisions is seen to be slightly larger than the same from smaller systems. At smaller  $p_T$ values, the $v_1/v_3$ falls sharply compared to $v_1/v_2$.

\begin{figure}[h]
\centerline{\includegraphics*[width=8.4 cm]{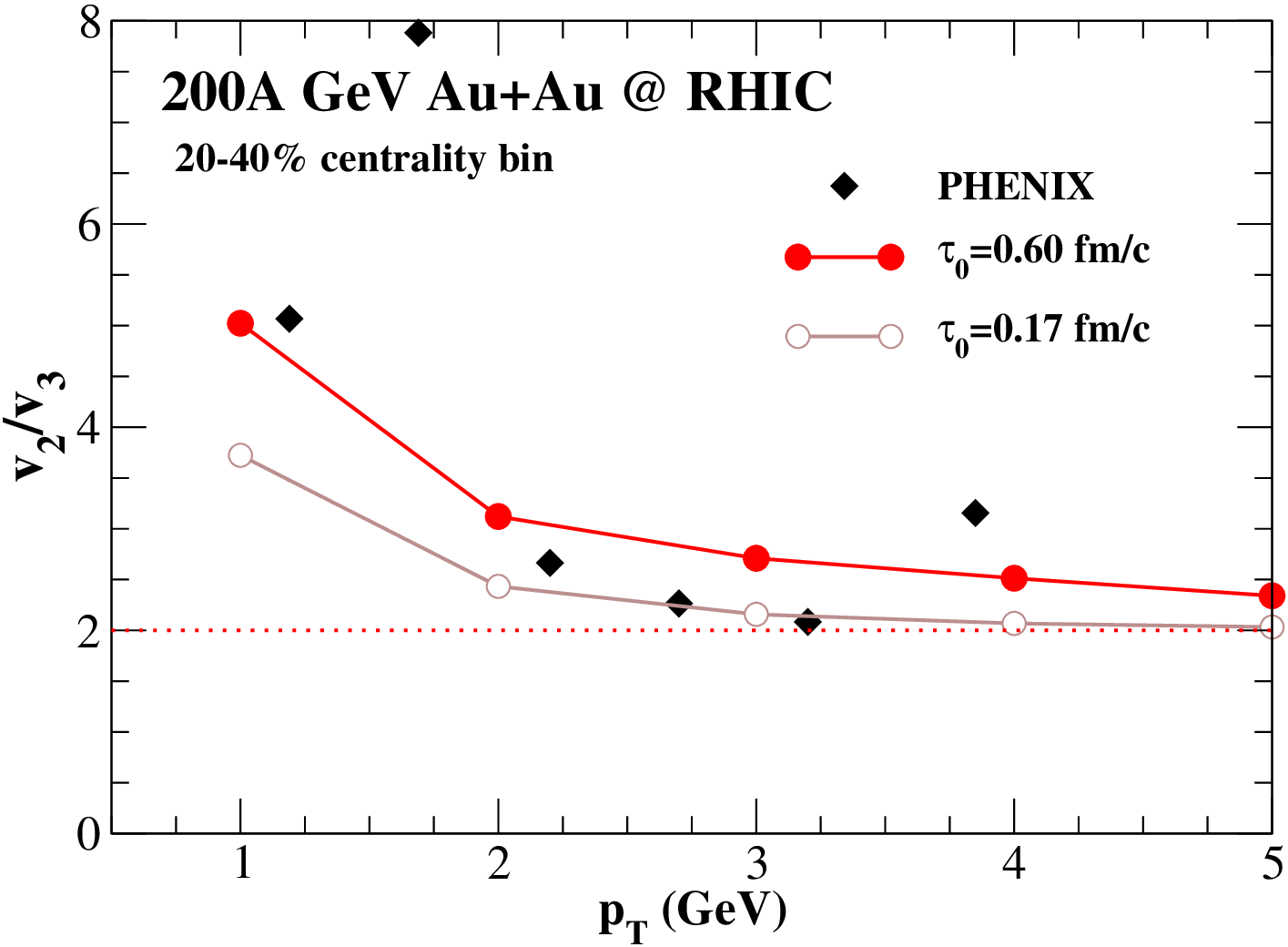}}
\caption{(Color online) The ratio of thermal photon $v_2$ and $v_3$ as a function of $p_T$  from Au+Au collisions at RHIC for 20--40\% centrality bin for  two different initial formation time $\tau_0$ as 0.17 fm/c and 0.60 fm/c. }
\label{ratio_tau}
\end{figure}

\begin{figure}[h]
\centerline{\includegraphics*[width=8.4 cm]{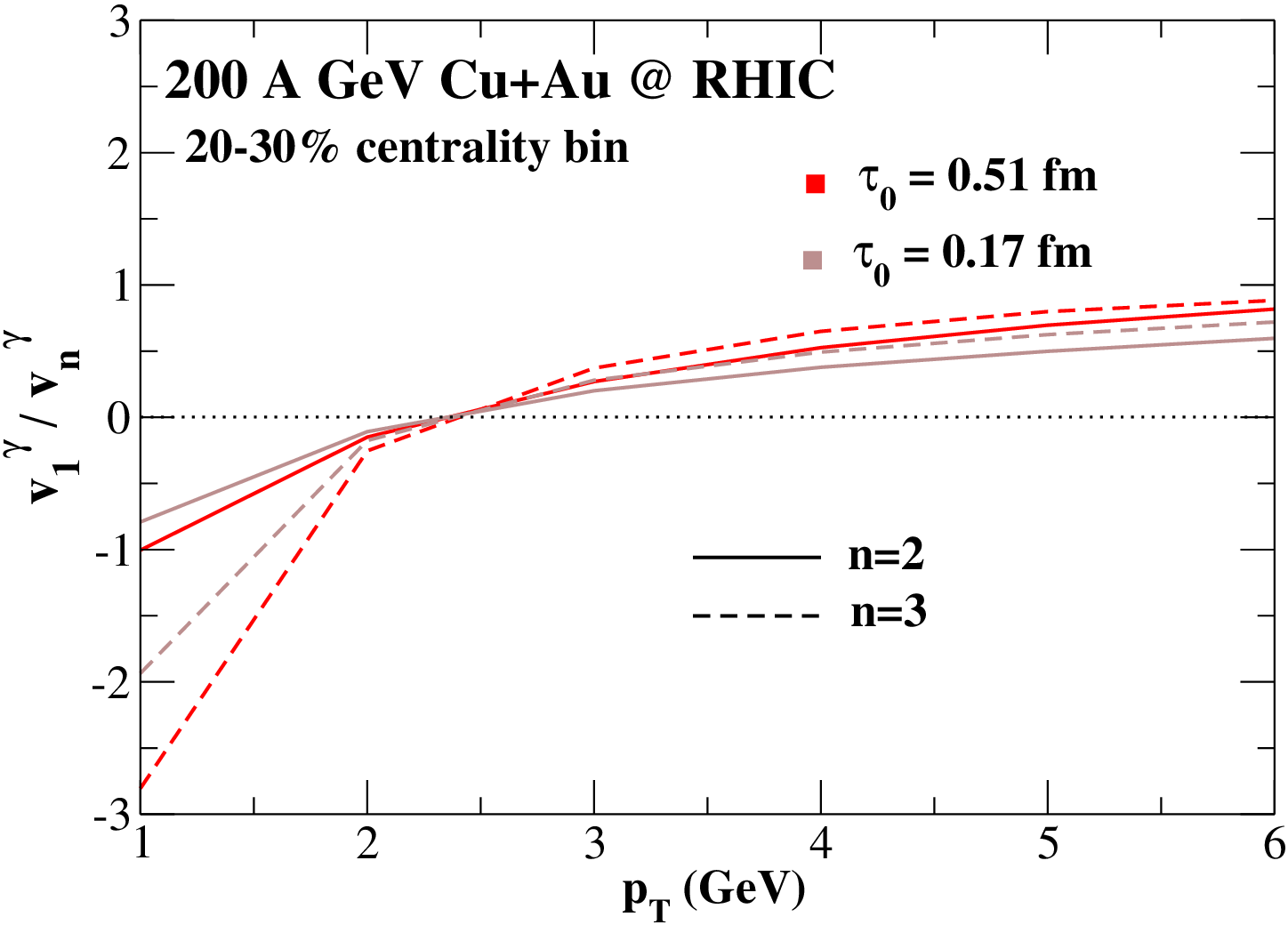}}
\caption{(Color online) The ratio of thermal photon $v_1$ and $v_2$  and  $v_1$ and $v_3$ as a function of $p_T$  from Cu+Au collisions at RHIC for 20--40\% centrality bin for  two different initial formation time $\tau_0$ as 0.17 fm/c and 0.51 fm/c. }
\label{ratio_tau_v1}
\end{figure}

\begin{figure}[h]
\centerline{\includegraphics*[width=8.4 cm]{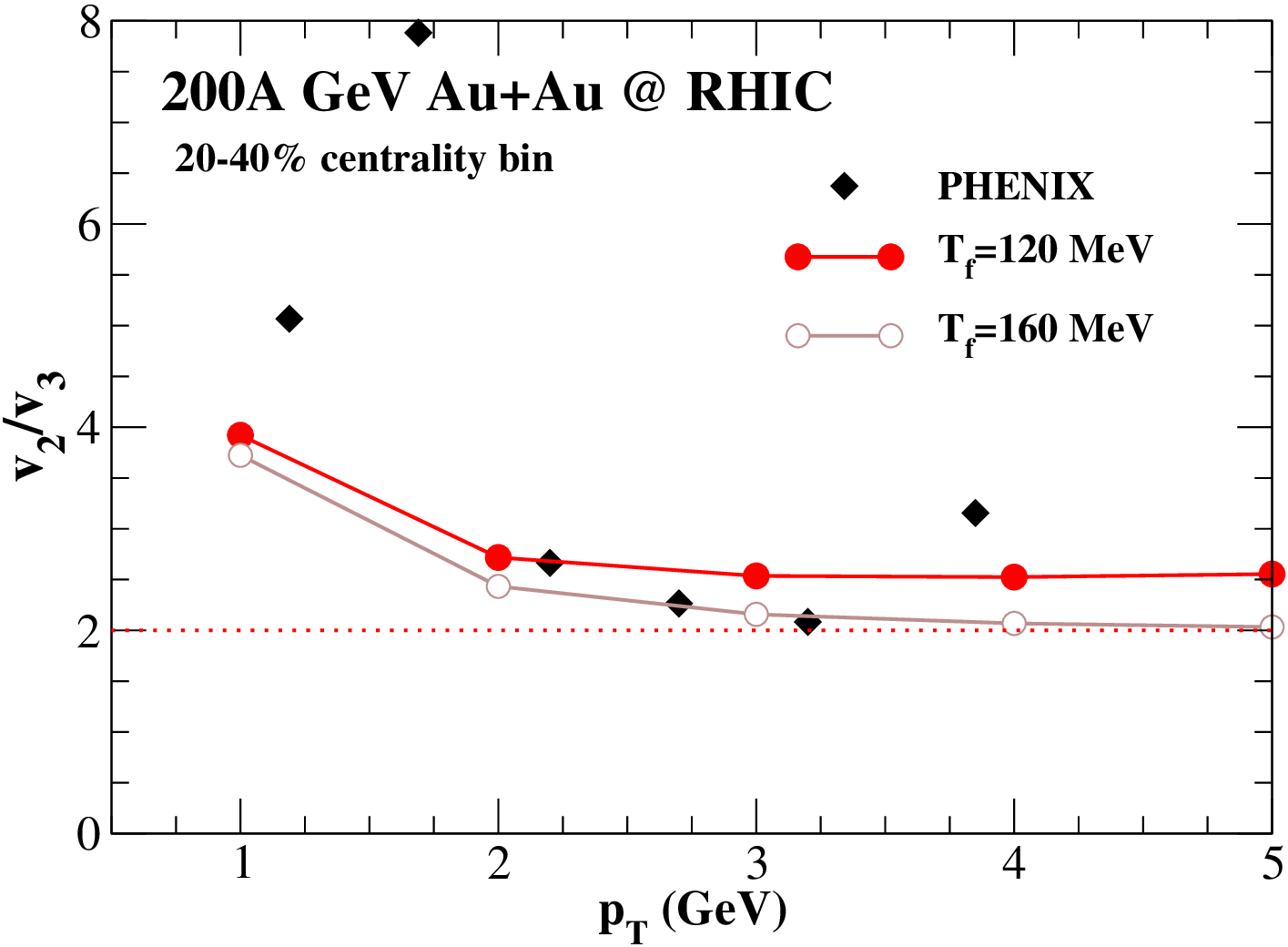}}
\caption{(Color online) The ratio of thermal photon $v_2$ and $v_3$ as a function of $p_T$ from Au+Au collisions at RHIC for 20--40\% centrality bin for two different thermal freeze-out temperatures $T_f$ as 160 MeV and 120 MeV.}
\label{ratio_tf}
\end{figure}


\section{Formation time and freeze-out conditions}
Earlier studies have shown that photon elliptic flow increases significantly in the region $p_T > $ 2 GeV when the formation time $\tau_0$ is changed from 0.2 fm/$c$ to 0.6 fm/$c$ in the hydrodynamical model calculation~\cite{rc_tau0,rc_uranium}. The triangular flow parameter of photons is also shows similar (to the elliptic flow parameter) $\tau_0$ dependent behaviour~\cite{rc_v3}.

The formation time at RHIC is taken from~\cite{ekrt} and we keep the value of $\tau_0$ fixed for all centrality bins of Au+Au collisions at RHIC. However, one can expect a larger $\tau_0$ for peripheral collisions as the system formed in those collisions will take longer time to thermalized compared to the central collisions~\cite{Chatterjee}. The estimation of the value of centrality dependent $\tau_0$ is not trivial and for the same reason we do not change the value of $\tau_0$  for peripheral collisions. Photon anisotropic flow  for the three different systems considered here are done using  value of $\tau_0$ 0.17 fm/$c$. One can also expect the value to be larger for smaller systems and lower beam energies. To check the sensitivities of the ratio  to the value of $\tau_0$ we calculate the $v_2/v_3$  considering a $\tau_0$ 0.6 fm/$c$ (keeping total entropy of the system fixed) and compare with the results obtained using 0.17 fm/$c$ for Au+Au collisions at RHIC (see Fig. 6).

 The change in $\tau_0$ increases the anisotropic flow at higher $p_T$. Whereas, the change in $\tau_0$ affects the ratio mostly in the lower $p_T$ region. Comparison with the experimental data shows that the ratio with smaller $\tau_0$ is closer to the data in the $p_T$ region 2--3.5 GeV.

The $v_1/v_3$ and $v_1/v_3$ results for two different $\tau_0$ (0.17 fm/$c$ and 0.51 fm/$c$) are shown in fig. 7.  The ratio with directed flow parameter is found to be less sensitive to the value of $\tau_0$ compared to $v_2/v_3$. For $p_T \ <$ 2 GeV only the $v_1/v_3$ results changes  significantly with change in the value of $\tau_0$.







We consider a constant freeze-out temperature  for all the systems at different centrality bins which is fixed by reproducing the charged particle multiplicity. However, it is to be noted that a smaller $T_f$ would enhance the contribution of the photons produced from the hadronic matter and the overall anisotropic flow parameter would also increase due to that. Thus, it is important to know the effect of changing freeze-out temperature on the ratio.

We consider a sufficiently smaller freeze-out temperature (120 MeV) and calculate the $v_n$ ratio. The results from two different $T_f$ values are shown in Fig.~\ref{ratio_tf}. One can see that even for a much smaller value of $T_f$, the ratio does not change much in the lower $p_T$ region and a small change can be observed only for  larger $p_T$. However, a smaller $T_f$ increases the photon $v_2$ and $v_3$ in the entire $p_T$ region. We have studied $v_1/v_n$ with different $T_f$ for C+Au  collisions at  RHIC~\cite{rc_cluster}, and we observe a significant change for $v_1/v_2$ with lower $T_f$. Such behaviour is expected as $v_2$ in contrary to $v_1$ and $v_3$ is very sensitive to the hadronic phase~\cite{rc_v1,rc_pramana}. Thus, $v_1/v_2$ can be a potential observable that explicitly sensitive to hadronic phase. 

The $v_2/v_3$ ratio  from Au+Au collisions is found to be close to the experimental data even with significantly different $\tau_0$ and $T_f$ values. We can say that the ratio is a relatively clean and robust quantity than the individual anisotropic flow parameters. 


\vspace{1.0 cm}

The results  presented here are calculated considering an ideal hydrodynamic model calculation.
It has been shown in Refs.~\cite{uli,Gale:2018ofa} that the effect of viscosity on the photon production and the anisotropic flow calculation is important. Inclusion of viscosity changes the energy momentum tensor  and also the distribution function. Detailed study have shown that inclusion of viscosity reduces the photon anisotropic flow  mostly in the larger $p_T$ region. 

It was shown in earlier studies that two different values of $\eta/s$ for MC Glauber and MC KLN initial conditions give same $v_2(p_T)$, however, the triangular flow parameter is found to be different for two initial conditions~\cite{visco}. As a results the $v_2/v_3$ ratio is also found to be significantly different for the two cases specially in the lower $p_T$ region. 
Thus ratio of photon anisotropic flow from experimental data would also be important to understand the viscosity effect as the result from inviscid model calculations does not show such strong sensitivity to the initial parameters.

\section{summary and conclusions}
We calculate  the ratio of elliptic and triangular flow of thermal photons as a function of $p_T$ at 200A GeV at RHIC from Au+Au, Cu+Au and Cu+Cu collisions. Although the individual elliptic and triangular flow parameters underpredict the PHENIX data points, the ratio from Au+Au collisions is found to be close to the  data  in the $p_T$ region 2 - 4 GeV which is believed to be dominated the by thermal radiation.
The ratio does not depend strongly on the initial parameters of the model calculation as we see the change in ratio is marginal when we increase the initial formation time of the plasma from 0.17 fm/$c$ to 0.60 fm/$c$ and also decrease the final freeze-out temperature from 160 MeV to 120 MeV. It is to be noted that the thermal photon spectra and anisotropic flow parameter changes significantly when $\tau_0$ is increased and $T_f$ is decreased in that range. 
The ration of photon $v_2$ (or $v_3$) with the directed flow parameter also shows interesting features as a function of $p_T$. From these results we conclude that the ratio of photon anisotropic flow can be a potential observable to understand the initial state as well as the direct photon puzzle better.



\section{Acknowledgements}  We would like to thank the kanaad and VECC  GRID computer facility. PD acknowledges the support of the National Natural Science Foundation of China under Grants No. 11835002 and No. 11961131011. Discussions with Dinesh Kumar Srivastava and Guo-Liang Ma are gratefully acknowledged.

\end{document}